\RequirePackage{fix-cm} 
\documentclass[aip,graphicx]{revtex4-1}
\pdfoutput=1
\usepackage{graphicx}
\usepackage{epstopdf}

\usepackage{multirow}
\usepackage{array}
\usepackage{amsmath,amssymb,mathrsfs}
\usepackage[T1]{fontenc}
\usepackage[utf8]{inputenc}

\usepackage{lipsum,caption,graphicx}
\usepackage{booktabs}
\usepackage{subfig}
\usepackage{caption,setspace}
\begin{document}
\vspace*{0.2in}

\begin{flushleft}
{\Large
  \textbf\newline{Fast simulations of highly-connected spiking cortical models using GPUs} 
}
\newline
\\
Bruno Golosio\textsuperscript{1,2*},
Gianmarco Tiddia\textsuperscript{1,2},
Chiara De Luca\textsuperscript{3,4},
Elena Pastorelli\textsuperscript{3,4},
Francesco Simula\textsuperscript{4},
Pier Stanislao Paolucci\textsuperscript{4},
\\

\bigskip
\textbf{1} Department of Physics, University of Cagliari, Italy\\
\textbf{2} Istituto Nazionale di Fisica Nucleare (INFN), Sezione di Cagliari, Italy\\
\textbf{3} Ph.D. Program in Behavioural Neuroscience, ``Sapienza'' University of Rome; Italy\\
\textbf{4} Istituto Nazionale di Fisica Nucleare (INFN), Sezione di Roma, Italy\\

\bigskip

* golosio@unica.it

\end{flushleft}
\section*{Abstract}
Over the past decade there has been a growing interest in the development of parallel hardware systems for simulating large-scale networks of spiking neurons.
Compared to other highly-parallel systems, GPU-accelerated solutions have the advantage of a relatively low cost and a great versatility, thanks also to the possibility of using the CUDA-C/C++ programming languages.
NeuronGPU is a GPU library for large-scale simulations of spiking neural network models, written in the C++ and CUDA-C++ programming languages, based on a novel spike-delivery algorithm. This library includes simple LIF (leaky-integrate-and-fire) neuron models as well as several multisynapse AdEx (adaptive-exponential-integrate-and-fire) neuron models with current or conductance based synapses, user definable models and different devices. The numerical solution of the differential equations of the dynamics of the AdEx models is performed through a parallel implementation, written in CUDA-C++, of the fifth-order Runge-Kutta  method with adaptive step-size control.
In this work we evaluate the performance of this library on the simulation of
a cortical microcircuit model, based on LIF neurons and current-based synapses, and on a balanced network of excitatory and inhibitory neurons, using AdEx neurons and conductance-based synapses.
On these models, we will show that the proposed library achieves state-of-the-art performance in terms of simulation time per second of biological activity. 
In particular, using a single NVIDIA GeForce RTX 2080 Ti GPU board, the full-scale cortical-microcircuit model, which includes about 77,000 neurons and $3 \cdot 10^8$ connections, can be simulated at a speed very close to real time,
while the simulation time of a balanced network of 1,000,000 AdEx neurons with 1,000 connections per neuron was about 70 s per second of biological activity.

{\bf Keywords:} Spiking neural network simulator, cortical microcircuits, adaptive exponential integrate-and-fire neuron model, conductance-based synapses, GPU 

\title{Fast simulations of highly-connected spiking cortical models using GPUs}

\section{Introduction}
The human brain is an extremely complex system, with a number of neurons in the order of 100 billions, an average number of connections per neuron in the order of 10 thousands, hundreds of different neuron types, several types of neurotransmitters and receptors. Because of this complexity, the simulation of brain activity at the level of signals produced by individual neurons is extremely demanding, even if it is limited to relatively small regions of the brain.
Therefore there is a growing interest in the development of high-performance hardware and software tools for efficient simulations of large-scale networks of spiking neuron models.
Some simulators, as for instance NEST \cite{Fardet_2020}, NEURON \cite{Carnevale_2006} and Brian \cite{BretteGoodman2009}, combine flexibility and simplicity of use with the possibility to simulate a wide range of spiking neuron and synaptic models. All three of these simulators offer support for multithread parallel computation for parallelization on a single computer and MPI support for distributed simulations on computer clusters.
On the other hand, a fertile field of research in recent decades has investigated the use of highly parallel hardware systems for simulating large-scale networks of spiking neurons. Such systems include custom made neuromorphic very-large-scale-integration (VLSI) circuits \cite{Indiveri_2011},  field programmable gate arrays (FPGAs) \cite{Wang_2018}  and systems based on graphical processing units (GPUs)
\cite{Brette_2012, sanders10, Garrido2011, Vitay2015, Chou2018, Knight2018}.
Compared to other highly-parallel systems, the latter have the advantage of a relatively low cost, a great versatility, thanks also to the possibility of using the CUDA (Compute Unified Device Architecture) platform, and a sustained technological development driven by the consumer market. General purpose computing on graphical processing units (GPGPU) is widely employed for massively parallel computing. GPGPUs can significantly reduce the processing time compared to multi-core CPU systems for tasks that require a high degree of parallelism, because a single GPU can perform thousands of core computations in parallel. However, in order to derive maximum benefit from GPGPU, the applications must be carefully designed taking into account the hardware architecture. CUDA is a parallel computing platform  and programming model that has been created by NVIDIA to allow software developers to take full advantage of the GPU capabilities \cite{sanders10}.
Over the past decade, several GPU-based spiking neural network simulators have been developed (see Brette and Goodman (2012) for a review). EDLUT \cite{Garrido2011} is a hybrid CPU/GPU spiking neural network simulator which combines time-driven (in GPU) and event-driven (in CPU) simulation methods to achieve real-time simulation of medium-size networks, which can be exploited in real-time experiments as for instance the control of a robotic arm. ANNarchy \cite{Vitay2015} is a simulator for distributed rate-coded or spiking neural networks, which provides a Python interface for the definition of the networks and generates optimized C++ code to actually run the simulation in parallel, using either OpenMP on CPU architectures or CUDA on GPUs. 
CARLsim \cite{Chou2018} is a GPU-accelerated library for simulating large-scale spiking neural network (SNN) based on the Izhikevich neuron  model, which provides a PyNN-like programming interface in C/C++.
Recently, the GeNN simulator \cite{Knight2018} achieved cutting edge performance in GPU-based simulation of spiking neural networks, achieving better performance than CPU-based clusters and neuromorphic systems in the simulation of the full-scale cortical microcircuit model proposed by Potjans and Diesmann
\cite{Potjans_2014}.
In this work we present a comprehensive GPU library for fast simulation of large-scale networks of spiking neurons, called NeuronGPU, which uses a novel GPU-optimized algorithm for spike delivery. This library can be used either in Python or in C/C++. The Python interface is very similar to that of the NEST simulator and allows interactive use of the library.
NeuronGPU was recently proposed for being integrated with the NEST neural simulator \cite{Golosio_2020}.
In the following sections, after a general description of the library and of the spike delivery algorithm, we will evaluate the library on two types of spiking neural network models:
\begin{itemize}
\item
  the Potjans-Diesmann cortical microcircuit model \cite{Potjans_2014}, based on the leaky-integrate-and-fire (LIF) neuron model, which describes the behavior of a region of the cerebral cortex having a surface of 1 $\text{mm}^2$ and
  includes about 77,000 neurons and $3 \cdot 10^8$ connections;
\item
  a balanced network of excitatory and inhibitory neurons \cite{Brunel2000},
  based on the adaptive-exponential-integrate-and-fire (AdEx) neuron model
  \cite{Brette_2005}, with up to 1,000,000 neurons and $10^9$
  connections.
\end{itemize}
We will show that, although the building time is larger compared to
other simulators, NeuronGPU achieves state-of-the-art performance
in terms of simulation time per unit time of biological activity.

\section{Materials and Methods}
\subsection{The NeuronGPU library}
NeuronGPU is a GPU library for simulation of large-scale networks of spiking neurons, written in the C++ and CUDA-C++ programming languages.
Currently it can simulate LIF models, different multisynapse AdEx models with current or conductance based synapses and two user definable neuron models.
The LIF model subthreshold dynamics is integrated by the {\it exact integration}
scheme described in \cite{Rotter_1999} on the time grid given by the simulation
time resolution.
On the other hand, the numerical solution of the differential equations of the AdEx dynamics is performed through a parallel implementation, written in CUDA C++, of the fifth-order Runge-Kutta  method with adaptive control of the step size
\cite{Press_1992}.
In general, NeuronGPU can simulate networks of any neuron model and synaptic current model whose dynamics can be described by a system of ordinary differential equations (ODEs).
The computations are carried out using mainly 32-bit floating point numerical precision, with the exception of some parts of the code for which double precision calculations are more appropriate, e.g. those in which a very large number of terms can be added.
Neuron parameters and connection weights and delays can be initialized either using fixed values or through arrays or probability distributions.
Neuron groups can be connected either using predefined connection rules (one-to-one, all-to-all, fixed indegree, fixed outdegree, fixed total number) or by user-defined connections.
In addition to the standard synapse model, nearest-neighbor spike-timing-dependent-plasticity (STDP) is also available \cite{Morrison2008}.
Different types of devices can be simulated, including Poisson signal generators, spike generators, multimeters and parrot neurons.
NeuronGPU includes an  efficient implementation of GPU-MPI communication among different nodes of a GPU cluster, however the performance of the proposed library on GPU clusters has not yet been thoroughly evaluated, therefore
this feature is not described in the present work.
The Python interface is very similar to that of NEST in main commands, use of dictionaries, connection rules, device and neuron model names and parameters.
The following Python code sample illustrates this strong similarity.
\begin{verbatim}
import neurongpu as ngpu
# create Poisson generator with rate poiss_rate
pg = ngpu.Create("poisson_generator")
poiss_rate = 12000.0
ngpu.SetStatus(pg, "rate", poiss_rate)
# Create n_neurons neurons with n_receptor receptor ports
# neuron model is multisynapse AdEx (aeif) with conductance-based synapse
# described by the beta function 
n_neurons = 10
n_receptor = 2
neuron = ngpu.Create("aeif_cond_beta", n_neurons, n_receptors)
# Initialize receptor parameters
E_rev = [0.0, -85.0]
tau_decay = [1.0, 1.0]
tau_rise = [1.0, 1.0]
ngpu.SetStatus(neuron, {"E_rev":E_rev, "tau_decay":tau_decay,
                        "tau_rise":tau_rise})
# Connect Poisson generator to neurons
poiss_weight = 0.05
poiss_delay = 2.0
conn_dict={"rule": "all_to_all"}
syn_dict={"weight": poiss_weight, "delay": poiss_delay, "receptor":0}
ngpu.Connect(poiss_gen, neuron, conn_dict, syn_dict)
\end{verbatim}
About 30 test scripts and C++ programs have been designed to check the correctness of neuron model dynamics, devices, spike delivery, connection rules.
Many of such tests use similar NEST simulations as reference.
Several examples in C++ and in Python are also available.
NeuronGPU is an open-source library, freely available on github from the web address \url{https://github.com/golosio/NeuronGPU}
under the terms of the GNU General Public License v3.0.

\subsection{The spike-delivery algorithm}
\begin{figure}
\centering
\includegraphics[width=\textwidth]{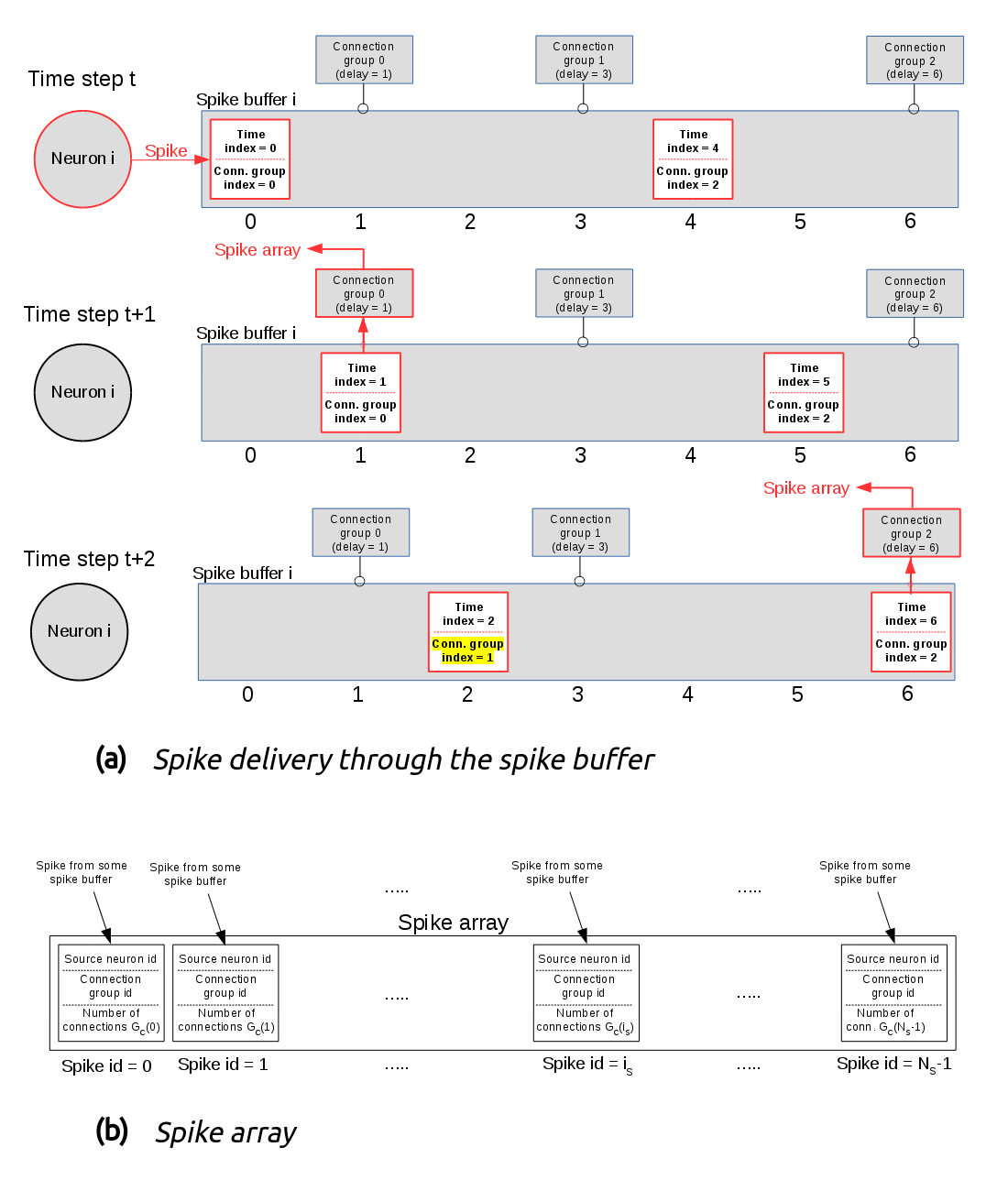} 
\caption{
a) Example of spike delivery through the spike buffer. At time  t,  the i-th neuron emits a spike which is inserted in the spike buffer. In this example, the buffer contains also another spike emitted previously. At each time step, the spike time index is incremented by 1. When it becomes equal to the delay of some connection group, the spike is delivered to that group and its connection group index is incremented by 1.
b) The spike array. When the time index of a spike matches the delay of a connection group, the spike is sent to the spike array, which is used for delivering the spike to all neurons of the connection group. 
}
\label{fig:spike:buffer}
\end{figure}

NeuronGPU uses one (output) spike buffer per neuron, which holds the spikes that have been fired by the neuron. The output connections of each neuron are organized in groups, all connection in the same group having the same delay (see Fig. \ref{fig:spike:buffer}). Only three values per spike are stored in the buffer: a multiplicity, a time index $t_s$, which starts from 0 and is incremented by 1 at every time step, and a connection-group index $i_g$, which also starts from zero and is incremented by 1 every time the spike reaches a connection group, i.e. when the time index $t_s$ matches the connection-group delay.
Figure \ref{fig:spike:buffer}(a) represents the structure of the spike buffer and illustrates an example of how the spike is delivered from the neuron that fired it to the target neurons of different connection groups. Keeping a connection-group index and having output-connection groups ordered according to their delays is useful for reducing the computational cost, because with this approach there is no need for a nested loop for comparing the time index of the spike with the connection delays. When the time index of a spike $t_s$ matches a connection-group delay, the spike is sent to the spike array, as shown in Fig. \ref{fig:spike:buffer}(b). Finally, spikes are sent from this array to the target neurons.
This final delivery is done directly by a CUDA kernel, so no additional memory is required. 
The maximum size of the global spike array is equal to the number of nodes (i.e. neurons and other spiking devices), so the maximum GPU memory required by this algorithm is well defined. 

In  MPI connections, when a source node (a neuron or another spiking device) is connected to target nodes of another host, a spike buffer, similar to the local one, is created in the remote host. When the source node fires a spike, this is sent to its spike buffer of the remote host, which delivers the spike to all target neurons after proper delays. 

\subsection{The Potjans-Diesmann cortical microcircuit model}
\begin{figure}
\centering
\includegraphics[width=0.5\textwidth]{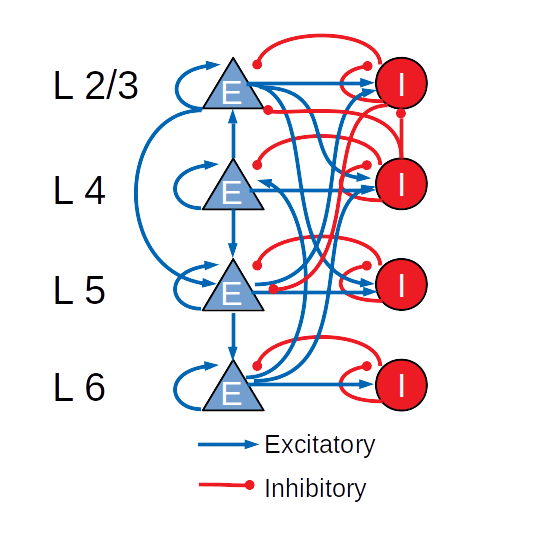} 
\caption{Schematic diagram of the Potjans-Diesmann cortical microcircuit model.}
\label{fig:potjans}
\end{figure}

The cortical microcircuit model used in this work was developed in 2014 by T. C. Potjans and M. Diesmann \cite{Potjans_2014} and describes a portion of
1 $\text{mm}^2$ of sensory cortex, comprising approximately 77,000 LIF neurons
organized into layers 2/3, 4, 5, and 6. Each layer contains an excitatory and an inhibitory population of LIF neurons with current-based synapses, for a total of 8 populations: 2/3I, 2/3E, 4I, 4E, 5I, 5E, 6I and 6E.
The number of neurons in each population, the connection probability matrix
and the rates of the external Poisson inputs are based on the integration of anatomical and physiological data mainly from cat V1 and rat S1.
The total number of connections is about $3 \cdot 10^8$.
Figure \ref{fig:potjans} shows a diagram of the model with a schematic representation
of the connections having probabilities > 0.04.

The LIF neuron model, used in the cortical microcircuit,
is one of the simplest spiking neuron models.
The neuron dynamics is modeled by the following
differential equation
\begin{equation}
  \tau_m \dfrac{dV_i}{dt} = -(V_i - E_{L}) + R_m I_{\text{syn},i}
\end{equation}
where $V_i(t)$ represents the membrane potential of neuron $i$ at time $t$,
$\tau_m$ is the membrane time constant, 
$E_L$ is the resting membrane potential,
$R_m$ is the membrane resistance
and $I_{\text{syn},i}$ is the synaptic input current.
In the exponential shaped postsynaptic currents (PSCs) model,
which will be used to simulate the Potjans-Diesmann cortical microcircuit model,
the input current is described by the following equation
\begin{equation}
  \tau_{\text{syn}} \dfrac{dI_{\text{syn},i}}{dt} = -I_{\text{syn},i}
  + \sum_i w_ {ij} \sum_{t_j^f} \delta (t - t_j^f)
\end{equation}
where $\tau_{\text{syn}}$ is the synaptic time constant,
$w_ {ij}$ are the connection weights
and $t_j^f$ are the spike times from presynaptic neuron $j$.

\subsection{The balanced network model}
\begin{figure}
\centering
\includegraphics[width=0.4\textwidth]{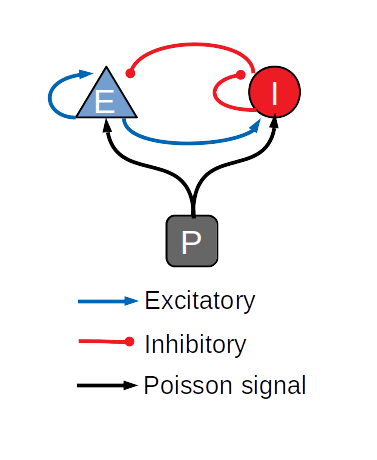} 
\caption{Schematic diagram of the balanced network used in the simulations.}
\label{fig:balanced_network}
\end{figure}

The performance of the library was also assessed on a balanced network of sparsely connected excitatory and inhibitory neurons \cite{Brunel2000},
using the AdEx neuron model with conductance-based synapses
and synaptic conductance modeled by an alpha function \cite{Roth_2013}.
Both populations of excitatory and inhibitory neurons are stimulated by an external Poissonian signal, as shown in Fig. \ref{fig:balanced_network}.
Simulations have been made with a variable number of neurons and connections,
with up to 1,000,000 neurons and $10^9$ connections.
Table \ref{tab:balanced_network} represents the paramenters used for the balanced network simulations.
\begin{table}
\centering
\begin{tabular}{ll}
\hline
\multicolumn{1}{c}{\textbf{Parameter}} & \multicolumn{1}{c}{\textbf{Value}} \\ \hline
$N_{ex}$ (n. of excitatory neurons) & variable \\
$N_{in}$ (n. of inhibitory neurons) & $N_{ex}/4$ \\
$CE$ (n. of input excitatory synapses per neuron) & variable \\
$CI$ (n. of input inhibitory synapses per neuron) & $CE/4$ \\
$W_{ex}$ (excitatory connection weight) & 0.05 \\
$W_{in}$ (inhibitory connection weight) & 0.35 \\
Mean delay & 0.5 ms \\
Delay STD & 0.25 ms \\
$W_{poisson}$ (Poisson signal weight) & 0.37 \\
$Rate_{poisson}$  (Poisson signal rate) & 20,000 Hz\\
Neuron average firing rate & 30.7 Hz\\ \hline
\end{tabular}
\caption{Values of the parameters used for the balanced network simulations.}
\label{tab:balanced_network}
\end{table}
The AdEx model represents an attractive neuron model for use in large-scale network simulations, because it is relatively simple compared to biologically detailed spiking neuron models, nonetheless it provides a good level of realism in representing the spiking behavior of biological neurons in many conditions, in the sense that it fits well the response of neurons as measured from electrophysiological recordings \cite{Brette_2005}. This model is described by a system of two differential equations. The first equation describes the dynamics of the membrane potential V(t) and includes an activation term with an exponential voltage dependence
\begin{equation}\label{eq:lif}
C\dfrac{dV}{dt}=-g_L(V-E_{L})+g_L\Delta_T e^{\frac{V-V_T}{\Delta_T}}+I_{syn}(V,t)-\omega+I_e
\end{equation}
where the synaptic current is
\begin{equation}
I_{\text{syn}}(V,t) = \sum_i g_i(t) (V - E_{\text{rev},i})
\end{equation}
$C$ is the membrane capacitance, $g_L$ is the leak conductance,
$E_L$ is the leak reversal potential, $\Delta T$ is a slope factor,
$V_T$ is the spike initiation threshold,
$\omega$ is the spike-adaptation current, $I_e$ is an external input current,
$g_i(t)$ are the synaptic conductances and $E_{\text{rev},i}$ are the reversal
potentials. The voltage is coupled to a second equation which describes
adaptation
\begin{equation}
\tau_{\omega}\dfrac{d\omega}{dt}=a(V-E_{L})-\omega
\end{equation}
where $\tau_{\omega}$ is the adaptation time-constant and $a$ is the
subthreshold adaptation parameter.  When the neuron fires a spike,
the adaptation current $\omega$ changes into
$\omega\rightarrow \omega +b$,
where $b$ is a spike-triggered adaptation parameter,
while the membrane potential changes into
$V\rightarrow V_r$.
Table \ref{tab:adex_params} reports the AdEx parameter values that have been used for the balanced network simulations. 

\begin{table}
\centering
\begin{tabular}{ll}
\hline
\multicolumn{1}{c}{\textbf{Parameter}} & \multicolumn{1}{c}{\textbf{Value}} \\ \hline
$C$ (Membrane capacitance) & 281 pF \\
$g_L$ (leak conductance) & 30 nS \\
$E_L$ (leak reversal potential) & -70.6 mV \\
$V_T$ (spike initiation threshold) & -50.4 mV \\
$\Delta T$  (slope factor) & 2 mV \\
$\tau _w$ (adaptation time constant) & 144 ms \\
$a$ (subthreshold adaptation) & 4 nS \\
$b$ (spike-trigger adaptation) & 80.5 pA \\
$V_r$ (reset value of $V_m$ after a spike) & -60 mV \\
$E_{ex}$ (excitatory reversal potential) & 0 mV \\
$E_{in}$  (inhibitory reversal potential) & -85 mV \\
$\tau_{\text{syn}}$ (synaptic time constant) & 1 ms \\ \hline
\end{tabular}
\caption{Values of the AdEx parameters used in the balanced network simulations.}
\label{tab:adex_params}
\end{table}

\section{Results}
The cortical microcircuit model and the balanced network described in the previous section were used both to verify the correctness of the simulations performed using NeuronGPU and to compare the performance of the proposed library with those of NEST version 2.20.0 and GeNN version 3.2.0.
For this purpose, we used a PC with a CPU Intel Core i9-9900K with a frequency of 3.6 GHz and 8 cores featuring hyperthreading with two threads per core, for a total number of 16 hardware threads, 64 GB RAM, and a GPU card NVIDIA GeForce RTX 2080 Ti with 11GB of GDDR6 VRAM, 4352 CUDA cores and a boost clock of 1635 MHz. NeuronGPU and GeNN simulations were also performed on a system equipped with an NVIDIA Tesla V100 GPU with 16 GB GPU memory and 5120 CUDA cores.

\subsection{Simulation of the cortical microcircuit model}
Following the procedure proposed by van Albada et al. \cite{vanAlbada2018} and by Knight and Nowotny \cite{Knight2018}, in this section we will verify the correctness of the simulations by comparing some relevant statistical distributions extracted from the simulations of the Potjans-Diesmann cortical microcircuit model made using NeuronGPU with the analogous distributions obtained using the NEST simulator.
Subsequently, still following the same line of \cite{vanAlbada2018} and \cite{Knight2018}, the cortical microcircuit model will be used as a benchmark to evaluate the performance of NeuronGPU in terms of building time and simulation time per unit time of biological activity.

The Python code used for simulations, available in
\url{https://github.com/golosio/NeuronGPU/tree/master/python/Potjans_2014}, is almost identical to the NEST implementation
(\url{https://nest-simulator.readthedocs.io/en/stable/microcircuit/}).
\begin{figure}[!htb]
\centering
\includegraphics[width=0.47\textwidth]{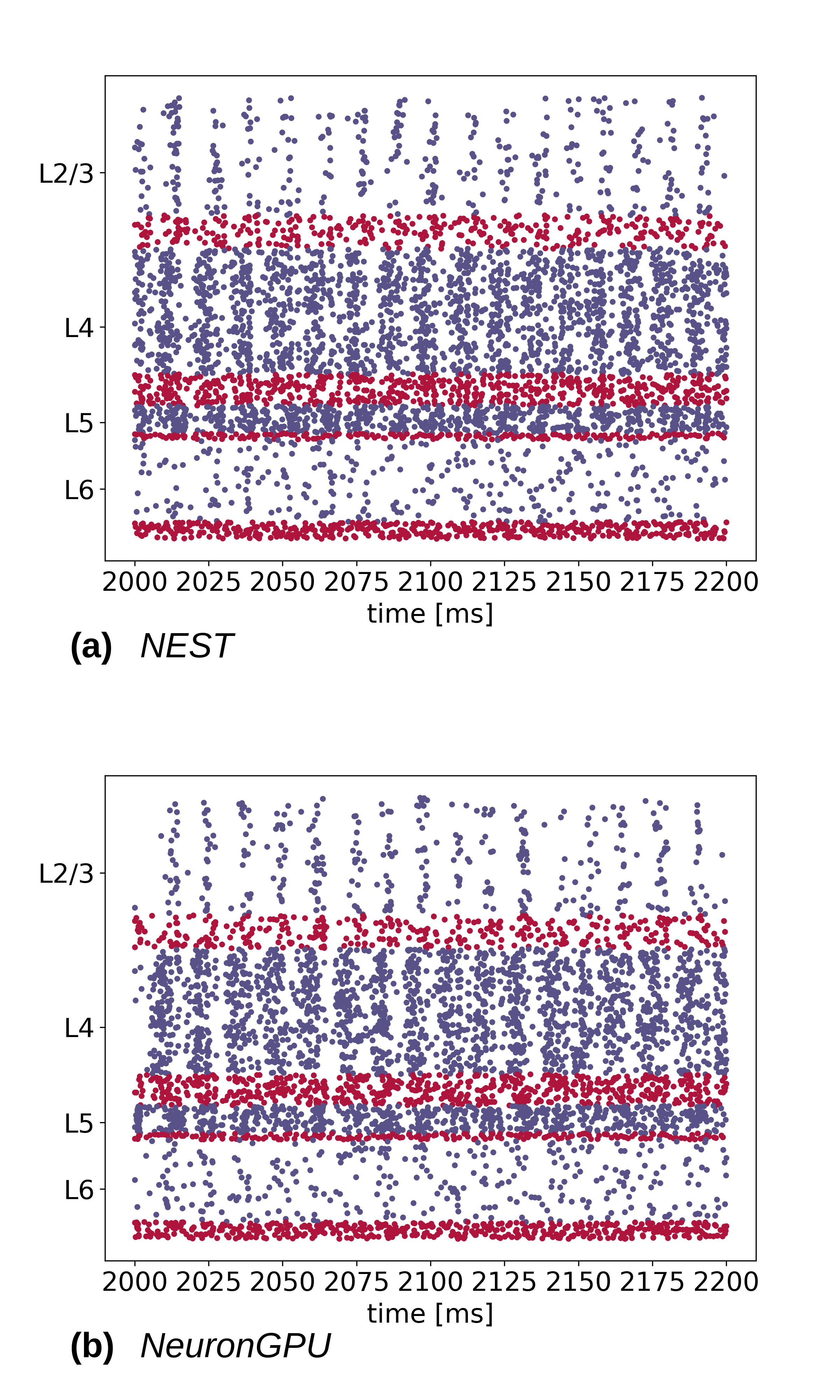}
\caption{Raster plot showing spike times (dots) of neurons from each population
  of the cortical microcircuit model, simulated using NEST and NeuronGPU, in a time window of 200 ms (in blue the excitatory and in red the inhibitory). Due to the high number of neurons in the model, only the spikes of one neuron out of ten are shown.}
\label{fig:raster_plot}
\end{figure}
Figure \ref{fig:raster_plot} shows a raster plot of the spike times
of neurons from each population of the model, simulated using NEST and
NeuronGPU, in a time window of 200 ms.

In order to verify the correctness of the simulations, we simulated 11 s of biological activity of the full-scale Potjans-Diesmann model with both NeuronGPU and NEST, with a time step of 0.1 ms.
For both simulators we performed 10 simulations, distinct from each other only for the initial seed for random number generation. As in ref. \cite{vanAlbada2018} and \cite{Knight2018}, the first second was discarded in order to eliminate transient trends.
The spike times of all neurons have been recorded during the simulations, and subsequently they have been used to extract three distributions for each population, namely:
\begin{itemize}
\item the average firing rate of the single neuron;
\item the coefficient of variation of the inter-spike time interval (CV ISI), defined as the ratio between the standard deviation and the average of the inter-spike time intervals;
\item the Pearson correlation between the spike trains.
\end{itemize}
The latter has been computed on a subset of 200 neurons for each population, as in ref. \cite{vanAlbada2018} and \cite{Knight2018}. This number represents a compromise between statistical precision and computation time. The spike trains of those neurons have first been rebinned to a time step of 2 ms,
equal to the refractory time.
Denoting the binned spike trains as $b_i$ and their mean value as $\mu_i$, the correlation coefficient between the spike trains $b_i$ and $b_j$ is defined as
$$C[i,j]=<b_i-\mu_i , b_j-\mu_j>/\sqrt{<b_i-\mu_i , b_i-\mu_i>\cdot<b_j-\mu_j , b_j-\mu_j>}$$
where $<,>$ represents the scalar product. For 200 spike trains, a 200x200 correlation matrix is returned.  The Pearson correlation distribution is evaluated as the distribution of the off-diagonal elements of this matrix.
All distributions have been evaluated from the spike time recordings using the Elephant (Electrophysiology Analysis Toolkit) package \cite{Elephant}, dedicated to the analysis of electrophysiological data in the Python environment.
The distributions have been smoothed using the KDE (Kernel Density Estimation) method \cite{Rosenblatt1956, Parzen1962},
available in the {\it sklearn} Python library through the function
\texttt{sklearn.neighbors.KernelDensity}.
The KDE method allows to estimate the probability density of a random variable with a reduced dependence on random fluctuations linked to individual simulations. In particular, each of the $N$ points belonging to a sample is represented by a Gaussian function of suitable width, called kernel bandwidth. The integral of each of these functions is normalized to $1 / N$; the overall distribution is therefore estimated as the sum of all these Gaussians, and obviously it has an integral normalized to one. The kernel bandwidth has been optimized using the so-called Silverman's rule \cite{silverman86}, which prescribes a bandwidth value of
\begin{equation}
  b = 0.9 \cdot \min\left(\hat{\sigma}, \frac{\text{IQR}}{1.349}\right)
  \cdot N^{-\frac{1}{5}}
\end{equation}
where $\hat{\sigma}$ is the standard deviation of the samples, $N$ is the sample size and IQR is the interquartile range.
It should be observed that the distributions obtained through the KDE method are continuous functions, since they are evaluated as the sum of a set of Gaussian functions.
\begin{figure}[h!]
\centering
\includegraphics[width=0.7\textwidth]{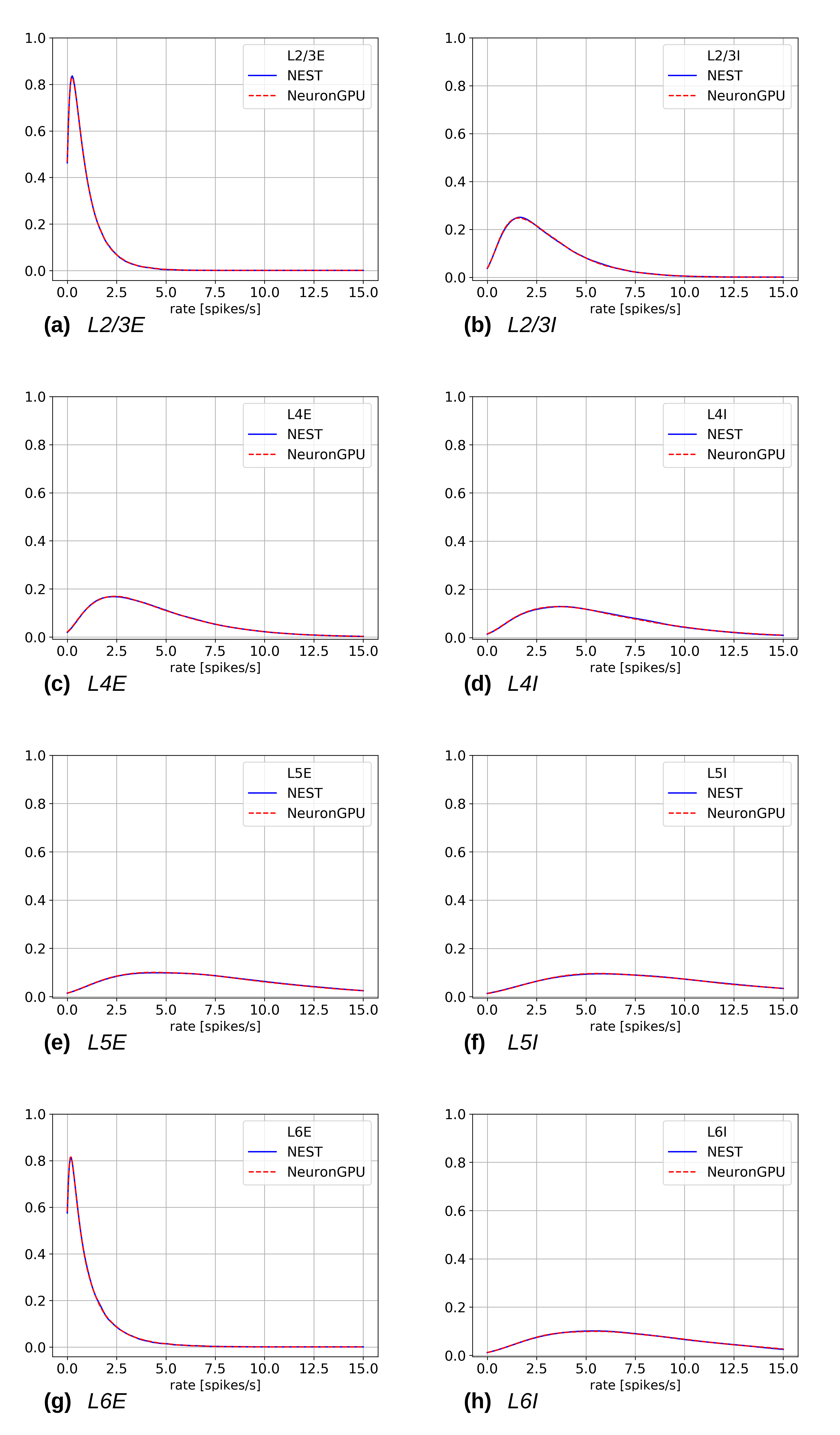}
\caption{Firing rate distributions for the eight populations of the cortical microcircuit model, averaged over 10 simulations made using NEST (blue) or NeuronGPU (red).}
\label{fig:firing_rate}
\end{figure}
\begin{figure}[h!]
\centering
\includegraphics[width=0.7\textwidth]{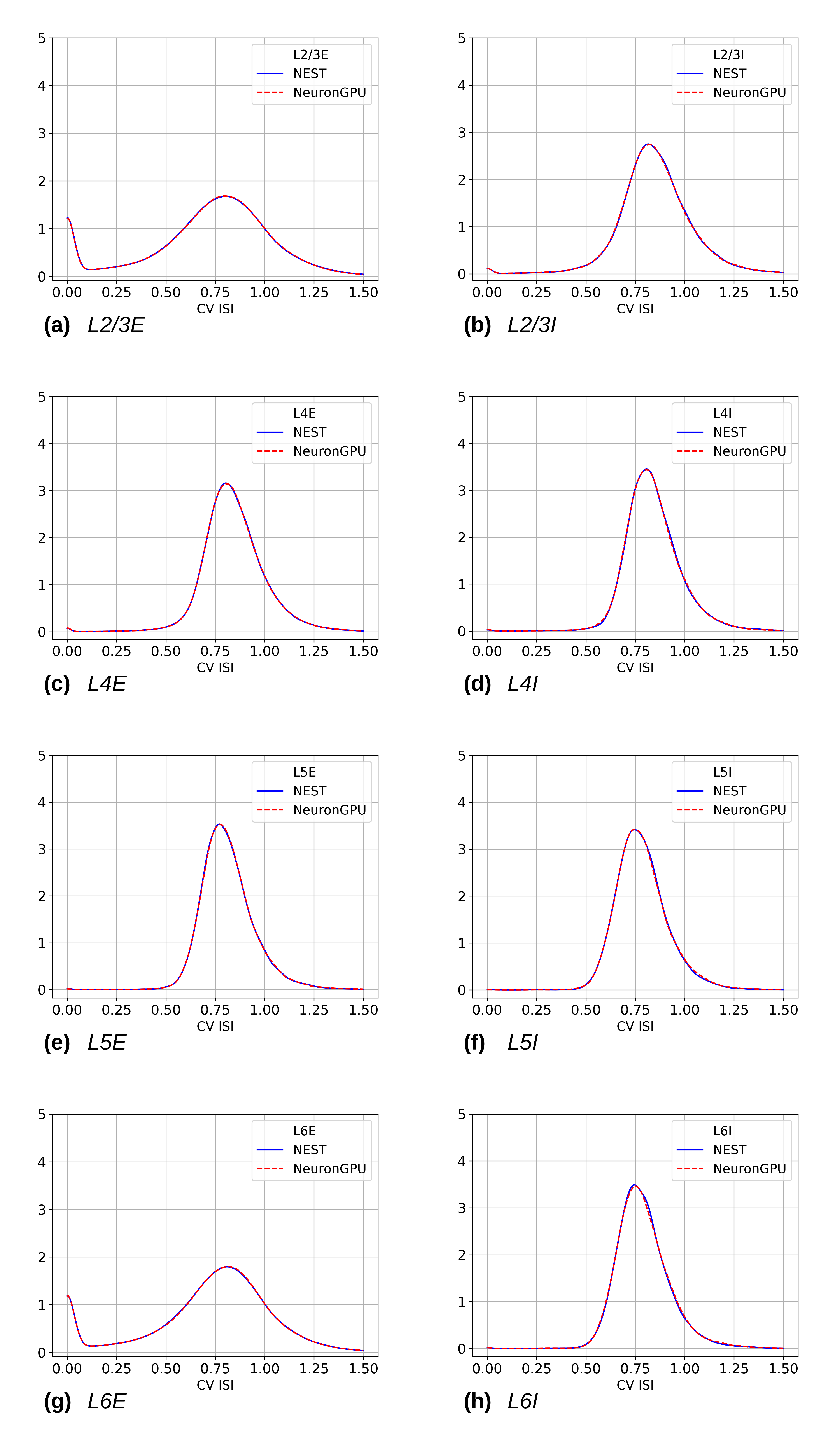}
\caption{Distribution of the coefficient of variation of interspike intervals (CV ISI) for the eight populations of the cortical microcircuit model, averaged over 10 simulations made using NEST (blue) or NeuronGPU (red).}
\label{fig:cv_isi}
\end{figure}
\begin{figure}[h!]
\centering
\includegraphics[width=0.7\textwidth]{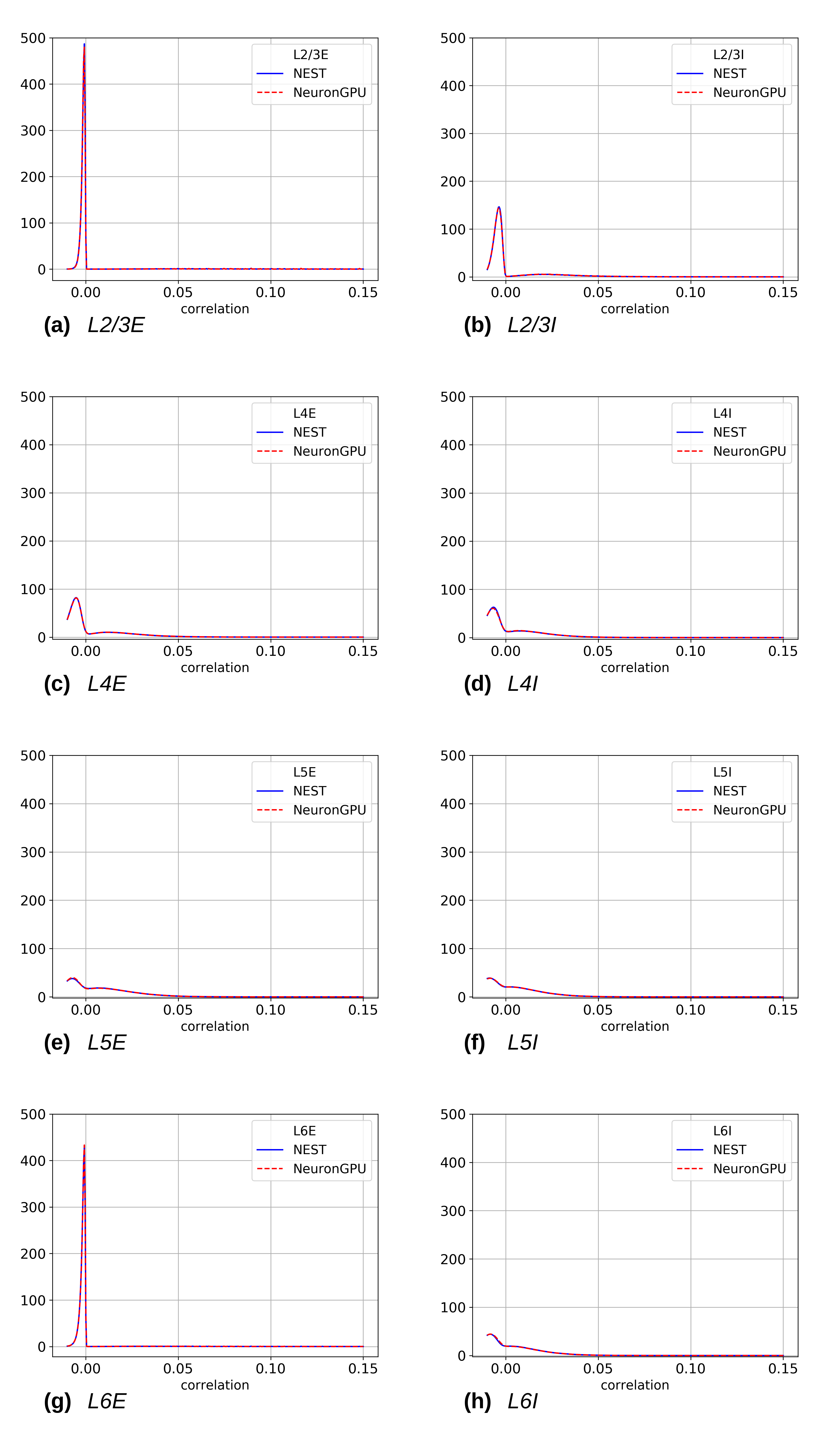}
\caption{Distribution of the Pearson correlation coefficient of the spike trains
 for the eight populations of the cortical microcircuit model, averaged over 10 simulations made using NEST (blue) or NeuronGPU (red).}
\label{fig:correl}
\end{figure}

Figures \ref{fig:firing_rate}, \ref{fig:cv_isi} and \ref{fig:correl}
show the distributions of the firing rate, the CV ISI and the Pearson correlation coefficient, respectively, for the 8 populations of the Potjans-Diesmann model, averaged over 10 simulations made using NEST or NeuronGPU.
As can be seen in the graphs, the distributions obtained from the two simulators are very similar to each other.
In order to compare the distributions obtained using NeuronGPU to those obtained using NEST, we evaluated the Kullback-Leibler (KL) divergence
\cite{Kullback1951}, defined as
$D_{KL}(p_1, p_2)=-\sum_i p_{1,i} \log(p_{1,i}/p_{2,i})$,
where $p_1$ and $p_2$ are two distributions, and the index $i$ runs on the sampling points of the two distributions.
For this purpose, we used 10 pairs of simulations (NeuronGPU-NEST and NEST-NEST) using different seeds for random number generation.
The KL divergence was then calculated for each pair and
its average and standard deviation were calculated on the 10 pairs.
Since the KDE method provides a smooth continuous function, the result is
not sensitive to the sampling step as long as this is small enough.
The KL divergence was evaluated using the Python scientific library and in particular the \texttt{scipy.stats.entropy} function.
Table \ref {tab:KL_divergence} shows the average and standard deviation of the KL divergences between the distributions of firing rates, CV ISI and Pearson correlation,
obtained from NEST and from  NeuronGPU simulations,
for the eight populations of the cortical microcircuit model.
It can be observed that the KL divergence between distributions obtained from NEST and from NeuronGPU are perfectly compatible with the divergence 
between distributions obtained from NEST simulations with different seeds.
\begin{table}[h!]
\centering
\resizebox{\textwidth}{!}{%
\begin{tabular}{@{}cccccccccccc@{}}
\multicolumn{12}{c}{$D_{KL}$ of the firing rate, CV ISI and Pearson correlation distributions}       \\ \midrule
& \multicolumn{3}{c} {$D_{KL}$ firing rate}
&& \multicolumn{3}{c} {$D_{KL}$ CV ISI}
&& \multicolumn{3}{c} {$D_{KL}$ Pearson correlation} \\
\cmidrule{2-4} \cmidrule{6-8} \cmidrule{10-12}
Population
& NEST, NeuronGPU && NEST, NEST
&& NEST, NeuronGPU && NEST, NEST
&& NEST, NeuronGPU && NEST, NEST
\\
\cmidrule{2-2} \cmidrule{4-4}
\cmidrule{6-6} \cmidrule{8-8}
\cmidrule{10-10} \cmidrule{12-12}
L2/3E                   & $0.0014\pm 0.0004$       && $0.0014 \pm 0.0003$
&& $0.0009\pm 0.0006$    && $0.0008 \pm 0.0002$
&& $0.037\pm 0.014$      && $0.037 \pm 0.014$ \\
L2/3I                   & $0.0020 \pm 0.0009$      && $0.0019 \pm 0.0007$
&& $0.003 \pm 0.001$     && $0.0034 \pm 0.0007$
&& $0.020 \pm 0.011$     && $0.017 \pm 0.005$ \\
L4E                     & $0.0004\pm 0.0001$       && $0.00047 \pm 0.00011$
&& $0.0012\pm 0.0003$    && $0.0013 \pm 0.0004$
&& $0.005 \pm 0.002$     && $0.005 \pm 0.002$ \\
L4I                     & $0.0012 \pm 0.0005$      && $0.0012 \pm 0.0003$
&& $0.0037 \pm0.0011$    && $0.0039 \pm 0.0015$
&& $0.005 \pm 0.002$     && $0.005 \pm 0.004$ \\
L5E                     & $0.0011 \pm 0.0005$      && $0.0011 \pm 0.0007$
&& $0.004 \pm 0.001$     && $0.0042 \pm 0.0013$
&& $0.005 \pm 0.005$     && $0.004 \pm 0.003$ \\
L5I                     & $0.003 \pm 0.002$        && $0.0029 \pm 0.0012$
&& $0.012 \pm 0.004$     && $0.012 \pm 0.002$
&& $0.004 \pm 0.002$     && $0.005 \pm 0.003$ \\
L6E                     & $0.0015 \pm 0.0003$      && $0.0014 \pm 0.0002$
&& $0.0011 \pm 0.0003$   && $0.0010 \pm 0.0003$
&& $0.023 \pm 0.003$     && $0.025 \pm 0.008$ \\
L6I                     & $0.0012 \pm 0.0005$      && $0.0014 \pm 0.0008$
&& $0.006 \pm 0.002$     && $0.0051 \pm 0.0008$
&& $0.008 \pm 0.006$     && $0.005 \pm 0.004$ \\
\bottomrule
\end{tabular}}
\caption{
  Kullback-Leibler divergence between the distributions of the firing rate, coefficient of variation of interspike intervals (CV ISI) and Pearson correlation coefficient, extracted from NEST and NeuronGPU simulations. The columns labeled as
{\it NEST-NeuronGPU} show the average values and the standard deviations of the divergence between NEST and NeuronGPU, while the columns labeled as
{\it NEST-NEST} show the same values for NEST simulations with different seeds.
}
\label{tab:KL_divergence}
\end{table}

To compare the performance of NeuronGPU with those of NEST and GeNN, we performed a series of 10 simulations of 10 s of biological activity of the cortical microcircuit with each simulator, using different seeds for random number generation.
The execution time of the simulations can be divided into building time and simulation time of biological activity.
The building time includes the time needed to allocate memory for neurons, devices and connections, to build connections and to initialize the values of state variables and parameters.
\begin{figure}[h!]
\centering
\includegraphics[width=0.8\textwidth]{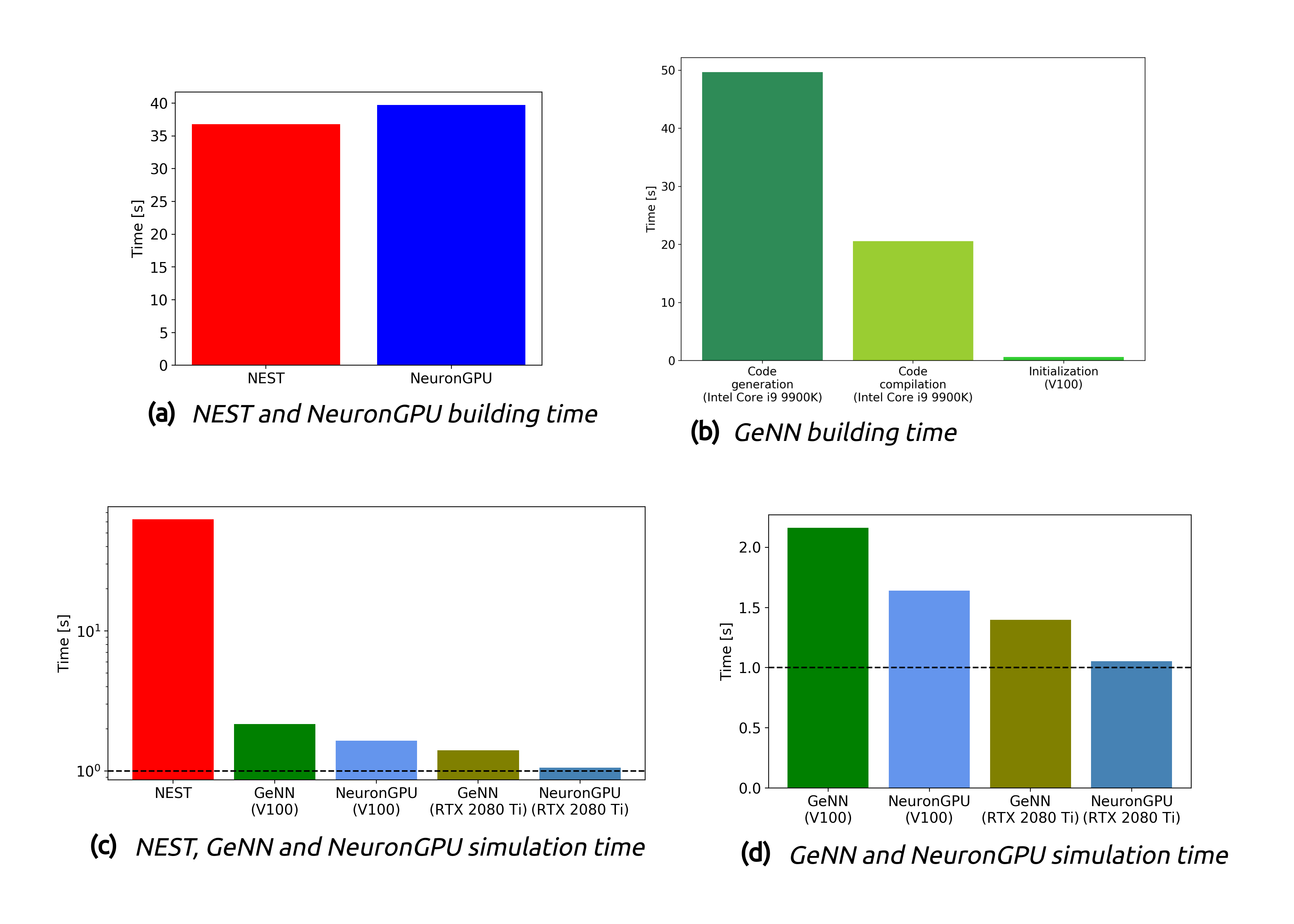}
\caption{(a) Building time of the cortical microcircuit model simulated with NEST and with NeuronGPU on a system equipped with an Intel Core i9-9900K CPU. (b) Times for code generation, compilation and initialization of the cortical microcircuit model for GeNN. The first two phases are performed by the CPU, and the times refer to a system equipped with an Intel Core i9-9900K. The third phase is mainly performed by the GPU, and the figure shows the time for an NVIDIA Tesla V100. (c),(d) Simulation times per second of biological time of the cortical microcircuit model simulated with NEST, NeuronGPU and GeNN on various CPU and GPU hardware. The horizontal line represents the biological time.}
\label{fig:microcircuit_perf}
\end{figure}
Figure \ref{fig:microcircuit_perf}(a) shows the building time for NEST and NeuronGPU.
On a system equipped with an Intel Core i9-9900K CPU,
the building times were $36.8 \pm 0.6$ s and $39.7 \pm 0.4$ s
for NEST and NeuronGPU respectively.

Figure \ref{fig:microcircuit_perf}(b) shows the time for code generation, compilation and initialization of the cortical microcircuit model for GeNN.

The building time of NeuronGPU is comparable to that of NEST.
This is due to the fact that in NeuronGPU the connections are initially created in the RAM, and only immediately before the simulation they are copied from RAM to GPU memory.
In GeNN the code of the model is generated from C/C++-like code fragments and it must be compiled before execution. Any changes in the model parameters require a new generation and compilation of the code. Once the code is generated and compiled, the initialization is very fast.

Figures \ref{fig:microcircuit_perf} (c) and (d) show the simulation times per unit time of biological activity for NeuronGPU, NEST and GeNN on different CPU and GPU platforms.
The simulation time per second of biological time with NEST running on the Intel Core i9-9900K CPU was $62.7 \pm 0.3$ s.
On a system equipped with an NVIDIA Tesla V100 GPU card, the simulation time per second of biological time with GeNN was 2.16 s. NeuronGPU was 31.6\% faster than GeNN, with a simulation time of
$1.641 \pm 0.014$ s on the same GPU.
On an NVIDIA RTX 2080 Ti GPU card, the simulation time per second of biological time with GeNN was $1.398 \pm 0.007$ s, while NeuronGPU was 32.5\% faster with a simulation time of $1.055 \pm 0.004$~s.

\subsection{Simulation of the balanced network model}

\begin{figure}[!htb]
\centering
\includegraphics[width=0.65\textwidth]{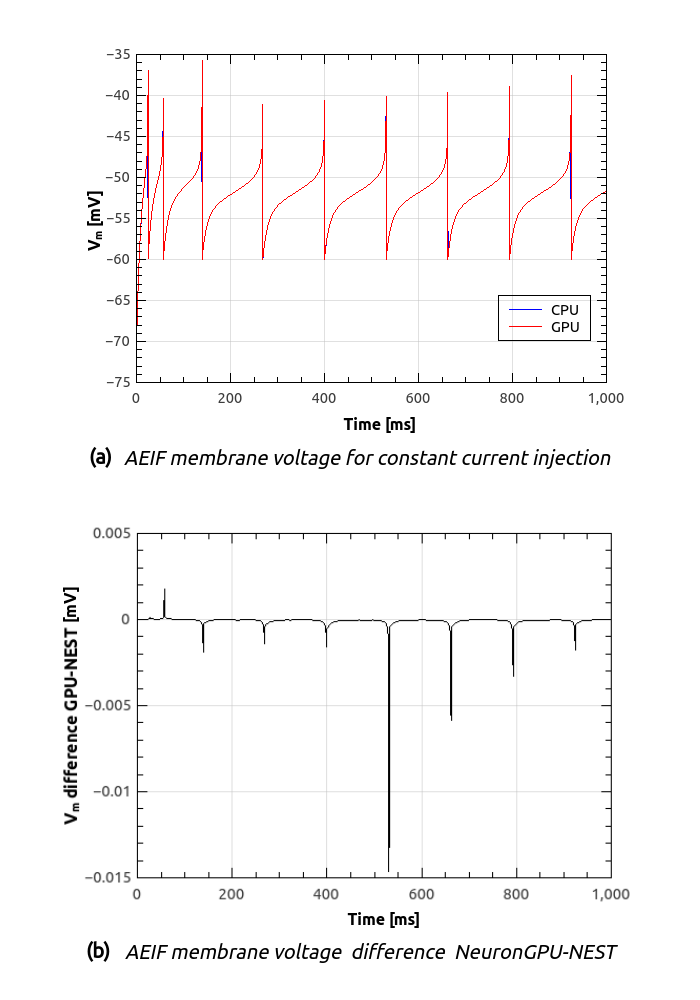}
\caption{Membrane voltage of an AdEx neuron with the parameter values reported in Table \ref{tab:adex_params} and an injected current of 700 pA, simulated with NeuronGPU and with NEST (a) and difference between the two simulation signals (b)}
\label{fig:constcurr}
\end{figure}
Figure \ref{fig:constcurr}(a) shows the time course of the membrane voltage of an AdEx neuron with the parameter values reported in Table \ref{tab:adex_params} and an injected current of 700 pA, simulated with NeuronGPU and with NEST. With the exception of the peaks, the two plots appear to be perfectly superimposed on this scale.
Figure \ref{fig:constcurr}(b) represents the difference between the two signals simulated with NEST and with NeuronGPU. Apart from the peaks, the difference is in the order of a few $10^{-4}$ mV.
\begin{figure}[!htb]
\centering
\includegraphics[width=0.5\textwidth]{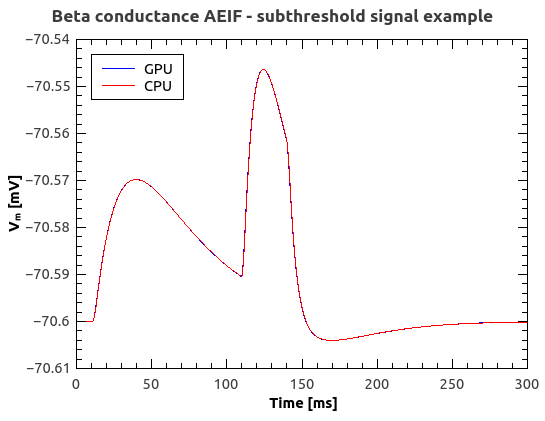} 
\caption{Membrane voltage of an AdEx neuron stimulated by three input spikes in a subthreshold condition, simulated using NeuronGPU and NEST.}
\label{fig:subthreshold}
\end{figure}
Figure \ref{fig:subthreshold} shows the time course of the membrane voltage of an AdEx neuron stimulated by three input spikes on three different receptor ports in a subthreshold condition, simulated with NeuronGPU and with NEST.

In the remaining part of this section we present the results of simulations of
the balanced network illustrated in Fig. \ref{fig:balanced_network},
with the parameters shown in Tables \ref{tab:balanced_network} and \ref{tab:adex_params}.
\begin{figure}[!htb]
\centering
\includegraphics[width=0.8\textwidth]{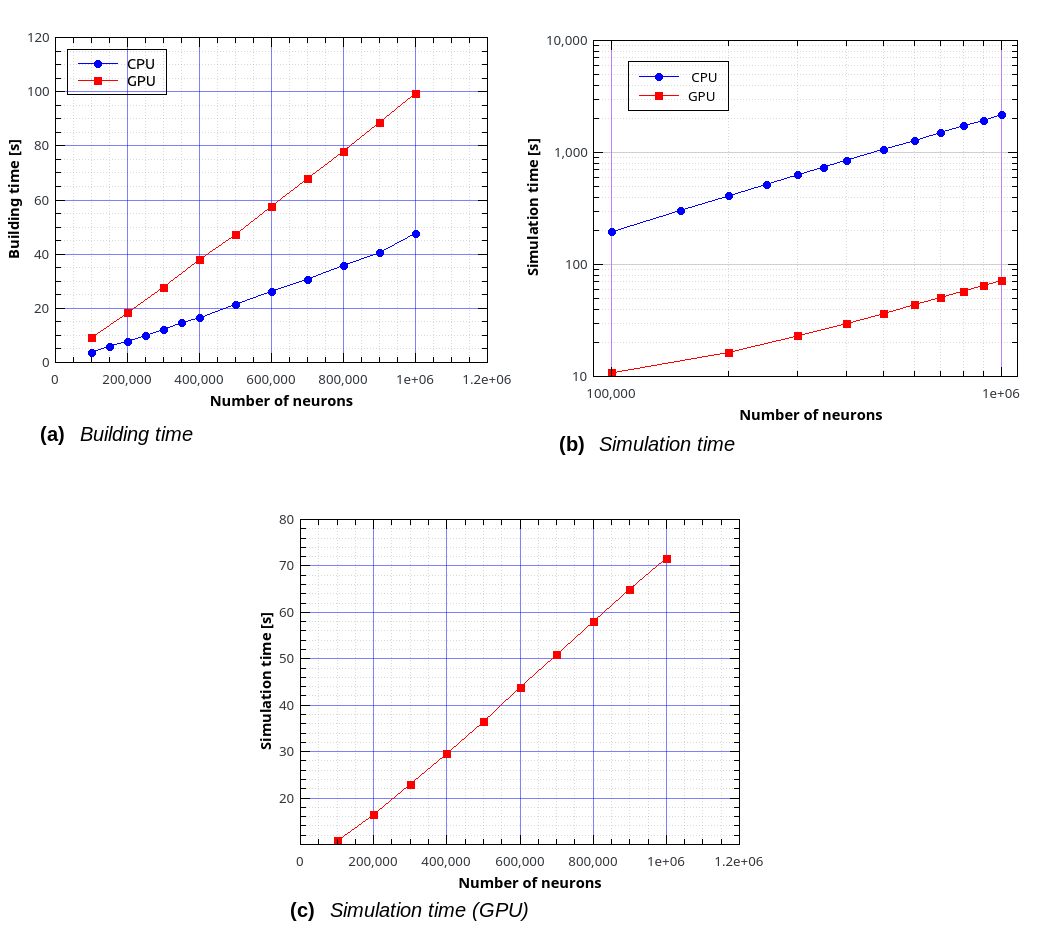}
\caption{
Building time (a) and simulation time (b) for the balanced network simulations with a variable number of neurons and a fixed number of 1000 input connections per neuron, simulated using NeuronGPU and NEST, and simulation time for NeuronGPU shown on a different scale (c).}
\label{fig:balanced_perf1}
\end{figure}
\begin{figure}[!htb] 
\centering
\includegraphics[width=0.8\textwidth]{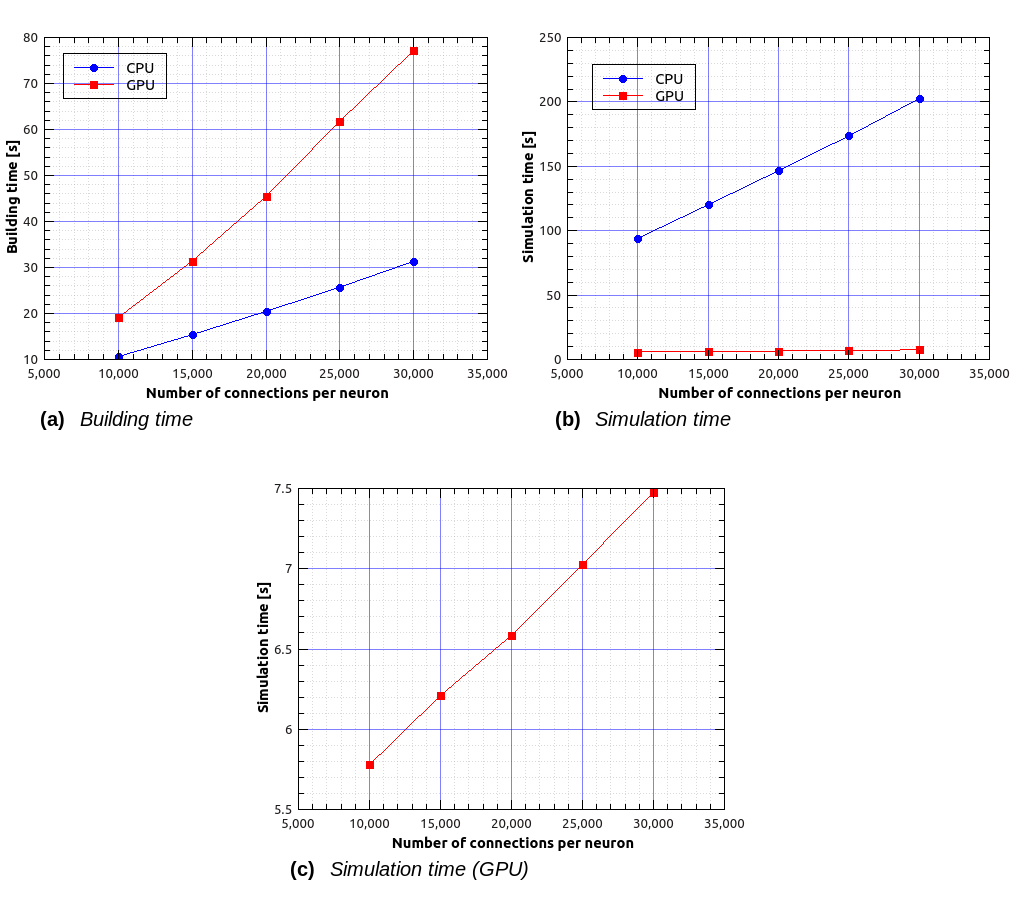}
\caption{
Building time (a) and simulation time (b) for the balanced network simulations with a fixed number of 30000 neurons and a variable number of input connections per neuron, simulated using NeuronGPU and NEST, and simulation time for NeuronGPU shown on a different scale (c).
}
\label{fig:balanced_perf2}
\end{figure}

Figure \ref{fig:balanced_perf1}(a) shows the building time for the balanced network simulations as a function of the number of neurons, for a fixed number of 1000 input connections per neuron.
Figure \ref{fig:balanced_perf1}(b) represents the simulation time per second of biological activity of the balanced network as a function of the total number of neurons. It can be observed that the GPU simulations are faster than the CPU’s by a factor ranging from about 18$\times$ for 100,000 neurons with $10^8$ connections
to 30.4$\times$ for $10^6$ neurons with $10^9$ connections.

Figure \ref{fig:balanced_perf2}(a) shows the building time as a function of the number of connections per neuron for a fixed total number of neurons, which was set to 30,000. Fig. \ref{fig:balanced_perf2}(b) represents the simulation time as a function of the number of connections per neuron. It can be observed that, in this case, simulations on GPU are faster than on CPU by a factor ranging from about 16$\times$ for 30,000 neurons with $3 \cdot 10^8$ connections to about 27$\times$ for 30,000 neurons with $9 \cdot 10^8$ connections.
 
\section{Discussion}

As it can be observed in figure \ref{fig:microcircuit_perf}, the building time of the cortical microcircuit model simulated using NeuronGPU is comparable to that of NEST, mainly because in NeuronGPU the connections are created in the RAM and only immediately before the simulation loop they are copied to the GPU memory.
Compared to most GPU-based simulators, NeuronGPU offers a wide range of choices for connection rules and connection parameter distributions, which can be exploited at runtime and interactively through the Python interface.
It is easier to manage these connection rules and these distributions on the CPU side, also thanks to the functions provided by the standard C++ library.
In both NEST and NeuronGPU the model parameters, the neuron populations and the network architecture are defined at runtime and the memory they need is allocated dynamically.
On the other hand, GeNN uses a code-generation approach.
The model parameters, neuron populations and architecture are defined using code fragments similar to C/C++, from which the CUDA/C ++ code of the model is generated. This code must be compiled before execution.
Any changes in the parameters, neuron populations or network architecture require a new generation and compilation of the code. Once the code is generated and compiled, the initialization is very fast as it is carried out directly by the GPU with parallel computing algorithms.
On the other hand, NeuronGPU achieved a simulation time per second of biological activity of 1.64 s on an NVIDIA Tesla V100 GPU 
and of 1.055 s on an NVIDIA RTX 2080 Ti GPU, about 32\% faster than GeNN, 59x faster than NEST and very close to biological time.

The building time of the balanced network simulated using NeuronGPU was about twice as large as that of NEST. However, NeuronGPU was faster than NEST in terms of simulation time per second of biological activity
by a factor ranging from about 16$\times$ for smaller networks to about 30$\times$
for networks with $10^9$ connections.
In future releases of the library, the building time could significantly be reduced by creating the connections directly in the GPU memory,
exploiting the parallel computing capabilities of the GPU and avoiding the bottleneck of memory transfer from RAM to GPU memory.
On the other hand, the high simulation speed demonstrated by the proposed library, significantly higher than that of other CPU and GPU based simulators, combined with the availability of a wide range of neuron models, devices and connection rules, makes this library particularly useful for simulations of large spiking neural networks over relatively long biological times. Furthermore, the high degree of similarity between the Python interfaces of NEST and NeuronGPU immediately simplifies porting scripts from one simulator to the other, and in the future opens the door to integration and cosimulations between NEST and NeuronGPU.

\section*{Conflict of Interest Statement}
%
The authors declare that the research was conducted in the absence of any commercial or financial relationships that could be construed as a potential conflict of interest.

\section*{Author Contributions}
%
BG, GT and PP wrote the manuscript. BG is the main developer of NeuronGPU. BG and PP designed the esperiments. All authors contributed to conducting the experiments and analyzing the results. All authors reviewed the manuscript.

\section*{Funding}
This work has been partially supported by the European Union Horizon 2020 Research and Innovation program under the FET Flagship Human Brain Project (grant agreement SGA3 n. 945539 and grant agreement SGA2 n. 785907) and by the INFN APE Parallel/Distributed Computing laboratory.
We acknowledge the use of Fenix Infrastructure resources, which are partially funded from the European Union's Horizon 2020 research and innovation programme through the ICEI project under the grant agreement No. 800858.

\section*{Acknowledgments}
We are grateful to Prof. Hans Ekkehard Plesser and to Dr. Tanguy Fardet for their revision of the aeif\_cond\_beta\_multisynapse model in the NEST simulator, which was the basis for the implementation of the same model in NeuronGPU. We would also like to thank Prof. Plesser, Prof. Markus Diesmann, Dr. Alexander Peyser and Dr. Wouter Klijn for the useful discussions on the dynamics of spiking neural networks, the use of CPU and GPU clusters and the spike delivery algorithms.

\section*{Data Availability Statement}
The source code of NeuronGPU can be downloaded from the web address \url{https://github.com/golosio/NeuronGPU}.
The Python code used for the cortical microcircuit simulations is available in
\url{https://github.com/golosio/NeuronGPU/tree/master/python/Potjans_2014}.
The Python code used for the balanced network simulation can be found in \url{https://github.com/golosio/NeuronGPU/blob/master/python/examples/brunel_net.py}.

\bibliographystyle{frontiersinSCNS_ENG_HUMS} 
\bibliography{neurongpu}


\end{document}